\documentclass[reprint,amsmath,amssymb, aps, prl]{revtex4-1}

\usepackage{graphicx}
\usepackage{dcolumn}
\usepackage{bm}
\usepackage{hyperref}

\usepackage{gensymb}

\begin{document}


\title{Evolution of black hole shadow in the presence of ultralight bosons}

\author{Rittick Roy}
 \email{rittickrr@gmail.com}
\author{Urjit A. Yajnik}%
 \email{yajnik@phy.iitb.ac.in}
\affiliation{%
 Department of Physics, Indian Institute of Technology Bombay, Mumbai 400076, India 
}%

\date{\today}

\begin{abstract}
Kerr black holes coupled to quantized bosonic fields display a special version of the Hawking effect, governed by the superradiance condition. This leads to rapid growth of boson cloud through spontaneous creation, leading to slowing down of the black hole, and detectable as growth of the black hole shadow. This can be developed into a technique for searching or constraining the existence of ultralight bosons. We study this phenomenon for spin-0 bosons in the shadow of a black hole, with a detailed analysis of Sgr$A^*$, and put estimates on the evolution time scales and subsequent change in the black hole shadow features. Our study shows that there is a small window of parameters where the increase of shadow of a supermassive black hole may be visible, but only if the sensitivity of measurements increases from current 25 $\mu$as to about 0.1 $\mu$as.
\end{abstract}

\maketitle

\section{Introduction}
\label{Intro}
Black holes have remained a laboratory of theoretical curiosities for the 100 years since their prediction. Their formation, stability and evolution have been  studied extensively \cite{Wald} and observational evidence in their favor has been mounting. The acceleration of extra-galactic cosmic rays \cite{Fermi:1949ee, Bhattacharjee:1998qc, Anchordoqui:2018qom}, production of highly relativistic jets in Blazars \cite{Halzen:2002pg} and the observed characteristics of Gamma-Ray Bursts (GRBs) \cite{Piran:1999kx} all show signatures of an inner core powered by a black hole. The detection of gravitational waves by the LIGO collaboration from the merging of black hole binaries \cite{Abbott:2016blz} in 2015 and subsequent events provided even stronger observational evidence in favor of black holes.

For photons in the black hole environment, there exists a least radius below which circular orbits are not sustainable resulting in collapse of the photons into the interior. The radius at which unstable circular orbit exists for photon is called the Photon Ring and for a Schwarzschild black hole lies at $r=3r_g$, where $r_g=GM/c^2$ is the gravitational radius of the black hole. It can be shown \cite{Weinberg:1972kfs}, that an incoming photon with an impact parameter $b<r_c=\sqrt{27}r_g$ would end up inside the photon ring and thus eventually inside the black hole, but the photons with $b>r_c$ will escape to infinity and create an `image' of the black hole that features a bright ring with a dark interior. The dark interior is evidence of the existence of a strong gravitational source and this image is commonly referred to as the `Shadow of a Black Hole'. A recent observational signature for black holes came from the imaging of the shadow of a supermassive black hole by the Event Horizon Telescope (EHT) collaboration \cite{Akiyama:2019cqa} which used a Very Long Baseline Interferometry (VLBI) array to observe the black hole at the centre of M87 (M$87^*$ from here onwards). Over the past decade, the excitement from the discovery of Sagittarius $A^*$ (Sgr $A^*$) at the centre of our galaxy has been growing, and is the next observational goal of the EHT. Such observations can give crucial signals to the near horizon dynamics of black holes \cite{Chen:2019fsq, Davoudiasl:2019nlo, Bambi:2019tjh, Vagnozzi:2019apd, Cunha:2019ikd}, one such signal that we study here is the development of ultralight boson clouds in the black hole environment.

The black hole shadow is subject to evolution due to the interaction of the black hole with its environment. The strong gravitational field of a Kerr black hole has been studied for its superradiant scattering effects. This happens when the rotational speed of spacetime itself exceeds that of the particle being scattered.  For waves the superradiance condition is that the frequency $\omega$ of an incoming wave satisfies \cite{Wald, Endlich:2016jgc}
\begin{equation}\label{eq:sr}
    0<\omega<m\omega_+
\end{equation}
were $\omega_+$ is the angular velocity of the black hole horizon and $m$ is the azimuthal angular momentum separation constant. This has been likened to Cerenkov effect and results in the outgoing flux being larger than the ingoing flux, at the cost of energy and angular momentum to the black hole. For extremely light particles the scattering solutions admit quantum field theoretic interpretation analogous to the Hawking radiation (and Schwinger effect), where quantum fluctuations can give rise to a virtual particle pair, of which one falls into the black hole and the other emerges (in Schwinger effect both emerge) as a net outgoing flux at infinity. Thus a quantised field of ultralight bosons coupled to Kerr geometry  gives rise to an intermediate effect, which has received attention in recent literature. In this, particles created by quantum fluctuations, and satisfying the superradiance condition do not emerge at infinity, but get bound in quasi-stationary states around the black hole. This process has been studied for both spin-0 \cite{Arvanitaki:2010sy} and spin-1 bosons \cite{Baryakhtar:2017ngi}, although we will focus on the spin-0 case in the current work. A precursor of this idea is the ``black hole bomb'' \cite{Press:1972zz, Hod:2014pza} wherein spontaneously created photons can destabilize the black hole when trapped in the black hole environment due to the presence of externally provided mirrors. As observed in \cite{Arvanitaki:2010sy} the occurrence of Hydrogen atom like bound states in Kerr geometry for ultralight bosons substitutes for the mirror like condition, and a black hole laser has been mooted in \cite{Landry:2016arq}. Such a phenomenon where particles arising from vacuum fluctuations remain confined in the vicinity of the strong gravitational source is best understood as \textit{quasi-Hawking} \textit{(qH)} effect.

Axions are a class of ultralight pseudo-scalar bosons proposed to resolve the $CP$ problem of strong interactions \cite{Peccei:2006as}, but occurring more generically in string theory, with possible values of rest mass ranging from $10^{-9}$ eV-$10^{-21}$ eV \cite{Arvanitaki:2010sy}. The bound states of these axions to black holes can result in the quasi-Hawking effect leading to a reduction of the spin parameter $a_*$ of the black hole which in turn could lead to substantial variation in the observed black hole shadow. In the current work we develop this idea to determine whether variation in the features of the black hole shadow is significant enough for detection and if so, whether it takes place over observational time scales. Based on this analysis, we identify the optimal candidate black hole for observation of this phenomenon. As mentioned before, we will focus on spin-0 bosons in this work which could be either scalar or pseudo-scalar. Hence, throughout the rest of the work `bosons' will refer to spin-0 bosons unless stated otherwise. We will use the $G=c=\hbar=1$ system of units for the rest of the work.  

\section{Boson cloud formation}
\label{bosonCloud}
Consider a Kerr black hole with mass $M$, angular momentum $J$ and scalar bosons of mass $m_s$. It is convenient to use the rescaled mass parameter $\mu_a=m_s G/\hbar c$, which in the system of units defined above expresses the mass of the scalar field. It has been shown \cite{Starobinsky:1973aij} that in the limit $\mu_aM\ll1$ and $\omega M \ll1 $, the Klein-Gordon equation satisfied by the bosons in the Kerr geometry takes a relatively simple form of the Schr\"odinger equation for Hydrogen atom with the fine structure constant replaced by the gravitational fine structure constant $\alpha=\mu_aM$.  The energy eigenvalues in this case develop imaginary parts \cite{Ternov, Detweiler:1980uk}, whereas the real part could be expressed in terms of the quantum number $n$ of the usual quantum numbers $nlm$ by 
\begin{equation}\label{eq:energy}
     \omega =\mu_a\Bigg(1-\frac{\alpha^2}{2n^2}\Bigg)
\end{equation}
On the other hand, the small imaginary part $\Gamma_{sr}$, gives rise to either damped or growing modes. In single particle quantum mechanics these modes may be interpreted either as trapped particles whose bound state is subject to eventual collapse, or whose amplitude grows, suggesting instability of the black hole \cite{Detweiler:1980uk}. However if we have ultra-light bosons, whose Compton wavelength is comparable to the black hole gravitational radius, the solutions may be interpreted to lead to spontaneous creation of particles and accumulation of the same in quasi-stationary modes around the black hole \cite{Ternov}. We refer to this as the quasi-Hawking effect.

During this process, bosons extract mass and angular momentum from the black hole and begin to populate the quasi-stationary states. This phenomenon leads to the formation of large boson clouds in the black hole environment. This black hole-boson cloud system is often referred to as the ``Gravitational Atom". 
The population of the cloud increases exponentially with time and the growth stops when enough spin has been extracted from the black hole such that the superradiance condition Eq. (\ref{eq:sr}) is no longer satisfied. The occupation number $N$ of a quantum level, denoted by $nlm$, that satisfies the superradiance condition will grow exponentially with time at a rate $\Gamma_{sr}$ \cite{Arvanitaki:2014wva}
\begin{equation}\label{eq:growth}
     \frac{dN}{dt}\Bigg|_{sr}=\Gamma_{sr}N
\end{equation}
After the process is complete, the maximum occupancy of a level with azimuthal quantum number $m$ could be expressed as \cite{Arvanitaki:2010sy}
 \begin{equation}\label{eq:Nmax}
     N_{max}\simeq 10^{76}\times \Bigg(\frac{\Delta a_*}{0.1}\Bigg)\Bigg(\frac{M}{10M_\odot}\Bigg)^2
 \end{equation}
where $\Delta a _* $ is the change in black hole spin. The entire process occurs over characteristic time scale $\tau_{sr}$ which depends on the mass of the black hole $M$, mass of the scalar field $\mu_a$ and the spin parameter $a_*$.

\section{Superradiance time scale}
\label{SupTime}
We define the parameter $j=\sqrt{l(l+1)}$ which denotes the angular momentum per boson in the boson-cloud. Within the approximation $\alpha/l\ll1$, it can be shown that for black hole masses $M_\odot- 10^{12} M_\odot$, the permitted range of values for the scalar field mass is $10^{-9}$ eV$\geq\mu_a\geq10^{-21}$ eV. The spin parameter of the black hole, at any point in time, could then be expressed in terms of $j$, $\mu_a$ and the occupation number $N$ as
\begin{equation}\label{eq:a_*}
    a_*(N)=\frac{J-Nj}{(M-N\mu_a)^2}
\end{equation}
where $J$ and $M$ are the initial angular momentum and mass of the black hole. Using this formula in Eq. (\ref{eq:growth}) we get the superradiant time scale expression as
\begin{equation}\label{eq:tsr}
    \tau_{sr}= \int_{a_{i}}^{a_{min}} \frac{1}{\Gamma_{sr}(a_*,\alpha)N}\frac{dN}{da_*} da_*
\end{equation}
where $a_{min}$ denotes the lowest spin of black hole before violating the superradiance condition, and $a_i$ is the initial black hole spin. Here $\frac{dN}{da_*}$ and $N(a_*)$ can be found by inverting Eq. (\ref{eq:a_*}). $a_{min}$ is determined by $\alpha$ through the superradiance condition Eq. (\ref{eq:sr})
\begin{equation}\label{eq:amin}
  a_{min}=\frac{2k_{nlm}}{1+k_{nlm}^2}
\end{equation}
where $k_{nlm}(\alpha)=(2\alpha/m)(1-\alpha^2/2n^2)$.  In the $\alpha/l\ll1$ limit, the superradiance rate $\Gamma_{sr}$ is well approximated by \cite{Arvanitaki:2014wva} 
\begin{equation}\label{eq:gamma}
   \Gamma_{sr} \approx \Gamma_{nlm}=2\mu_a\alpha^{4l+4}r_+(m\omega_+-\mu_a)C_{nlm}
\end{equation}
where 
\begin{align}\label{eq:Cnlm}
&C_{nlm}=\frac{2^{4l+2}(l+n)!}{n^{2l+4}(n-l-1)!}\Bigg(\frac{l!}{(2l)!(2l+1)!}\Bigg)^2\nonumber\\ &\times \prod_{i=1}^{l}\Bigg(i^2(1-a_*^2)+4r_+^2(m\omega_+-\mu_a)^2\Bigg)
\end{align}

where $r_+=M+\sqrt{M^2-a^2}$ represents the event horizon of the black hole. The sign of $\Gamma_{sr}$ is determined by the sign of the term $(m\omega_+-\mu_a)$, in agreement with Eq. (\ref{eq:sr}) in the low $\alpha/l$ limit. 

As shown in \cite{Arvanitaki:2010sy}, the function $\Gamma_{sr}$ drops exponentially with an increase in the quantum number $l$. Hence the bosons would start filling the states  corresponding to different values of $l$, with $l=1$ being the fastest growing level. Due to the subsequent slow down of the black hole, this will also be the first level to fall below the requirement of the superradiance condition. The subsequent evolution has many possibilities, and if observable can refine the confirmation of this process, but they do not interfere with the effect we discuss and we shall focus on the main effect. We will be concerned with the $l=1$ level from here onwards, unless mentioned otherwise. 

Given a value for the initial spin of the black hole $a_i$, we can put an upper limit on the desirable value of $\alpha$ since the minimum spin should be less than the initial spin of the black hole i.e. $a_{min}(\alpha)< a_i$. 
Taking a near extremal spin of $a_i=0.99$ and the condition $\alpha\ll1$, we get the upper bound on $\alpha$ to be $0.42$. 
$\tau_{sr}$ is found to be relatively insensitive to the possible $a_i$ values. However, our numerical study shows that $\tau_{sr}$ has an $\sim\alpha^{-5.7}$ dependence and hence increases sharply for lower values of $\alpha$. From numerical integration, it is found out that $\alpha>0.19$ is a reasonable estimate to keep the time scale in observational limits, viz., $\leq75$ years for $10^5M_\odot$. Thus the optimistic range of values of $\alpha$ is $0.19\leq\alpha\leq0.42$. Since $\alpha \sim M m_s$, we can still have a large possible range for the mass of the bosons by choosing different mass for black holes.
Choosing a value of $\alpha=0.28$ and $a_i=0.99$, we find the superradiance time scale as
\begin{equation}\label{eq:tsrfinal}
    \tau_{sr}\approx 5.43\times \left(\frac{M}{10^5M_\odot}\right)\left(\frac{0.28}{\alpha}\right)^{5.7} \  \mathrm{years}
\end{equation}
It should be noted here that the dependence of $\tau_{sr}$ on $\alpha^{-5.7}$ is an approximation while the dependence on $M$ is exact. The precise value of the superradiant time scales can be obtained by numerically integrating Eq. (\ref{eq:tsr}). The values of the superradiance times reported in this paper are calculated by numerical integration of Eq. (\ref{eq:tsr}).

\section{Simulation and prediction}
\label{sim}
We have simulated the shadow of a Kerr black hole  for a static observer at infinity using the contour equations as provided in \cite{Bardeen:1973tla, Cunha:2018acu}. The features of the shadow contour of a black hole could be expressed in terms of the `Shadow Parameter' $\delta x$ defined as
\begin{equation}\label{eq:shadowpar}
    \delta x=(x_{max}-x_{min})
\end{equation}
and its `Circularity' defined as 
\begin{equation}\label{eq:circ}
    \chi=\frac{(y_{max}-y_{min})}{(x_{max}-x_{min})}
\end{equation}
where $(x,y)$ denotes the coordinates of the black hole contour in the local sky of an observer at infinity. The parameter $\delta x$ and $\chi$ depends on the spin parameter and inclination $\theta_0$, which is the angle between the line of sight and the axis of rotation of the black hole. Table \ref{tab:table1} shows the shadow parameter and the fractional change in mass ($\delta m= (M_i-M_f)/M_i$), due to the extraction of mass by the boson cloud, from our simulation during different stages of the quasi-Hawking process. Note that the net extracted mass $\Delta M=(M_i-M_f)$ is calculated by multiplying the maximum occupancy of the boson cloud with the mass of individual bosons i.e. $\Delta M=N_{max}\mu_a$, where $N_{max}$ is given by Eq. (\ref{eq:Nmax}).
\begin{table}[b]
\caption{\label{tab:table1}%
The shadow parameter for an observer at infinity, with $\theta_0$= $90^\circ$, $60^\circ$, $17^\circ$ has been given for five values of the spin parameter $a_*$ corresponding to five values of the $\alpha$ parameter: $0.43$, $0.36$, $0.28$, $0.24$, $0.19$. The fractional change in mass of the black hole is also given in terms of $\delta m=\frac{\Delta M}{M_i}$ for an initial spin of $a_i=0.995$. The shadow parameters are given in units of the gravitational radius $r_g$
}
\begin{ruledtabular}
\begin{tabular}{ c c c c c }
\textrm{$a_*$}&
\textrm{$100\times\delta m$}&
\textrm{$\delta x\ (90^\circ)$}&
\textrm{$\delta x\ (60^\circ)$}&
\textrm{$\delta x\ (17^\circ)$}\\
\colrule
0.995 & 0 & 9.27 & 9.29 & 9.60\\ 
0.95 & 2.21 & 9.48 & 9.49 & 9.70\\ 
0.85 & 4.33 & 9.75 & 9.76 & 9.87\\ 
0.78 & 4.98 & 9.89 & 9.90 & 9.97\\ 
0.66 & 5.41 & 10.08 & 10.08 & 10.12\\ 
\end{tabular}
\end{ruledtabular}
\end{table}
\begin{table}[hbt]
\caption{\label{tab:table2}%
Taking the mass of Sgr$A^*$ as $M=4.2\times 10^6M_\odot$ and inclination angle $\theta_0=60^\circ$, we present the corresponding quantities namely $a_{min}$, $\tau_{sr}$, $m_s$, $\Delta d$ and $\Delta \chi$ for five representative values of $\alpha$ and for a fixed initial spin parameter $a_i=0.995$ ($\chi_i=0.902$):
}
\begin{ruledtabular}
\begin{tabular}{c c c c c c }
\textrm{$\alpha$}&
\textrm{$a_{min}$}&
\textrm{$\tau_{sr} (years)$}&
\textrm{$m_s (eV)$}&
\textrm{$\Delta d (\mu as) $}&
\textrm{$\Delta \chi$}\\
\colrule
$0.19$ &  $0.662$ & $2369.43$ & $3.81\times 10^{-17}$ & $1.30$ & 0.077 \\ 
$0.24 $& $ 0.777 $& $509.15 $& $4.81\times 10^{-17} $&$ 0.71$ & 0.060 \\ 
$0.28 $&  $0.848 $& $216.55 $& $5.62\times 10^{-17} $& $0.32$ & 0.046\\
$0.32 $&  $0.903$ & $121.61 $& $6.42\times 10^{-17} $& $0.07$ & 0.034\\
$0.34 $&  $0.924$ & $100.51 $& $6.82\times 10^{-17} $& $0.01$ & 0.028\\
\end{tabular}
\end{ruledtabular}
\end{table}
The recent imaging of the shadow of the black hole at the centre of M87 shows a bright ring $(\sim 42\pm 3\ \mu as)$ and a dark region at the centre with a contrast depression of about $10:1$ with the ring \cite{Akiyama:2019cqa}. This is evidence of the presence of an extremely compact object that lenses light from the accretion disk to create a shadow. The angular diameter of the ring is determined by the formula 
\begin{equation}\label{eq:ringdia}
    d=42\times\Bigg(\frac{\delta x_i}{11}\Bigg)\Bigg(\frac{M}{6.5\times10^9M_\odot}\Bigg)\Bigg( \frac{16.8\ Mpc}{D}\Bigg) \mu as
\end{equation}
where $D$ is the distance from Earth to the black hole. Therefore a change in the black hole ring diameter could be given by 
\begin{equation}\label{eq:delringdia}
    \Delta d=d\times\Bigg[\Bigg(\frac{\delta x_f}{\delta x_i}\Bigg)(1-\delta m)-1\Bigg]\ \mu as
\end{equation}
where $\delta x_i$ and $\delta x_f$ are the initial and final shadow parameters before and after the quasi-Hawking process. For a change in the spin parameter $\Delta a_*= 0.1$ (from $0.95$ to $0.85$ for $\theta_0=17^\circ$ in Table \ref{tab:table1}) and mass $\delta m=0.043$  the angular diameter of the ring of M$87^*$, with an initial mass of $\sim 6.2\times 10^9 M_\odot$ and a distance of $\sim 16.8$ Mpc \cite{Akiyama:2019cqa}, changes by $\Delta d \sim 0.28\ \mu$as. For observations at a wavelength of 1.3 mm, the EHT collaboration has a theoretical diffraction-limit resolution of about $\sim25\ \mu$as \cite{Akiyama:2019cqa}, making this variation far below the resolution of EHT. Moreover, the excessive mass stretches this variation over a duration of $4.5\times10^{5}$ years. Hence, the observation of shadow evolution of M$87^*$ and similar higher mass black holes will remain unrealistic in this scenario.

A more promising case is the supermassive black hole at the centre of our galaxy, Sagittarius $A^*$ under observation by the EHT collaboration
for which  $M$ lies in the range $\sim 3.3\times 10^6 M_\odot$ \cite{Aschenbach:2006vb} to  $4.1\times 10^6 M_\odot$  \cite{Ghez:2008ms, Gillessen:2008qv}, $\theta_0=60^\circ$ \cite{Roelofs:2019nmh} and $a_*$ is determined to be $\sim 0.99676$ \cite{Aschenbach:2004fx, Aschenbach:2006vb}.  Table \ref{tab:table2} shows the time scales of variation, the change in the shadow diameter $\Delta d$ and change in circularity of the shadow from our simulations for different values of the $\alpha$ parameter. For observation to be made, the time scale must be as low as possible with the change in shadow being as large as possible. Therefore there exists a tradeoff since $\tau_{sr}$, $\Delta d$ and $\Delta \chi$ all increases with an increase in $\alpha$. As already mentioned, with a resolution limit of 25 $\mu$as, the variations in the shadow feature of Sgr$A^*$ is far below the resolution of EHT. Improvements in the resolution of EHT has already been suggested with the possibility that EHT analyses can attain the accuracy of $1.5\ \mu$as \cite{Johannsen:2015hib} as well as the proposed imaging techniques at lower wavelengths $\sim 0.8$ mm and with space-based interferometers aiming to achieve a resolution of $\sim3\ \mu$as \cite{Fish}. As is evident from Table \ref{tab:table2}, the search for ultralight bosons in the range $\alpha=(0.3-0.2)$ might become instrumental in the future of black hole imaging with resolution in the order of 0.1 $\mu$as.    

The ratio of $M$ and $D$ must satisfy $M/D\geq1.64\times10^{16}$ kg/m for the ring diameter $d$ to be large enough for observation in the current EHT resolution. If we focus our attention to black holes with mass $M\leq10^{6}M_\odot$ which have favourable superradiant time scales, 
we can put an upper bound on $D$ given by $D\leq3.95\times 10^3$ pc. This is the intra-galactic scale and hence we only need to focus on black holes inside the Milky Way galaxy. 
For a solar mass black hole with $M=10M_\odot$ inside the milky way, $D$ must be $3.95\times10^{-2}$ pc for its ring diameter to be observable. The nearest such black hole candidate is the V616 Monocerotis which has $D\sim10^3$ pc \cite{Cantrell:2010vh} from Earth thus violating the requirement on $M/D$.  
For the present the only candidates which satisfy the $M/D$ constraint are M$87^*$ and Sgr$A^*$ and since the time duration of the effect for M$87^*$ is way too large for observations, the optimal candidate for observation of shadow evolution is Sgr$A^*$. An important assumption of our analysis lies in the fact that there should be no previous population of boson cloud in the black hole environment. But even if the boson cloud levels are previously populated, the population depletes through self annihilation as discussed in \cite{Arvanitaki:2014wva}. This depletion process takes place over $\tau_{regge}\sim 10^{11}$ yr for black holes of $M\sim10^6M_\odot$ for the $l=1$ level \cite{Arvanitaki:2014wva}. And only after depletion of lower $l$ levels, the higher $l$ levels could start to fill. Thus, the effect might also be observed for older black holes, by observing shadow evolution for higher $l$ levels, the essential analysis for which remains the same as done in this work.  

\section{Conclusion}
\label{conclusion}
We have searched for signatures of Ultralight Boson cloud evolution in the features of a black hole shadow. Our analysis shows the variation in the shadow due to boson cloud evolution is far below the current EHT resolution, even for the most optimistic case. But such searches may become instrumental in future EHT with the optimal candidate for shadow evolution being Sgr$A^*$ subject to different parameters such as its spin and mass of the ultralight bosons. Further, the discoverability of such an effect would improve significantly if  a population of $10^3 M_\odot-10^5 M_\odot$ black holes were to be found inside the Milky Way galaxy.

Another interesting signature of Ultralight Boson Cloud from black hole shadow can be obtained if the backreaction of the boson cloud becomes significant enough on the background metric. The photon geodesics in the black hole environment would then get perturbed and hence the shape of the shadow will change. This change will be reflected in the features of the shadow of the black hole. A very close study in this respect using backward ray tracing has been done in \cite{Cunha:2015yba}. 

In conclusion, the next best thing to observing the Hawking effect itself would be to observe quasi-Hawking effect obtaining in the vicinity of Kerr black holes where the superradiance condition can be satisfied. We have shown that the shadow of a black hole could be used to detect such an effect, simultaneously providing a detection mechanism for such ultra-light bosonic particles.

\section{Acknowledgement}
\label{ack}
We would like to thank M. B. Paranjape for useful discussion and comments.

\nocite{*}






\end{document}